\documentclass[conference]{IEEEtran}
\IEEEoverridecommandlockouts
\usepackage{xcolor,soul,framed} 
\colorlet{shadecolor}{yellow}
\usepackage[pdftex]{graphicx}
\graphicspath{{../pdf/}{../jpeg/}}
\DeclareGraphicsExtensions{.pdf,.jpeg,.png}
\usepackage[cmex10]{amsmath}
\usepackage{array}
\usepackage{ulem}

\usepackage{mdwmath}
\usepackage{mdwtab}
\usepackage{eqparbox}
\usepackage{url}
\usepackage{booktabs}
\usepackage{multirow}
\usepackage{cite}
\usepackage{algorithmic, algorithm}
\usepackage{graphicx}
\usepackage{subfigure}
\usepackage{caption}
\renewcommand{\emph}[1]{\textit{#1}}
\usepackage{amsmath,amssymb,amsfonts}
\newtheorem{theorem}{$\mathbf{Theorem}$}
\newtheorem{lemma}[theorem]{$\mathbf{Lemma}$}
\DeclareMathOperator{\Tr}{Tr}

\usepackage[a4paper, top=0.75in, bottom=1.3in, left=0.6in, right=0.6in]{geometry}

\begin{document}
\title{ CSI-Free Low-Complexity Remote State Estimation over Wireless MIMO Fading Channels using Semantic Analog Aggregation}

\author{\IEEEauthorblockN{Minjie~Tang$^*$,  Photios A. Stavrou$^*$, and Marios Kountouris${^*}^{\dagger}$}\\
$^*$Communication Systems Department, EURECOM, Sophia-Antipolis, France\\
$^{\dagger}$Department of Computer Science and
Artificial Intelligence, University of Granada, Spain\\
Emails: \texttt{\{Minjie.Tang, Fotios.Stavrou, marios.kountouris\}@eurecom.fr}}

\maketitle
\thispagestyle{empty}
\pagestyle{empty}

\begin{abstract}
In this work, we investigate low-complexity remote system state estimation over wireless multiple-input-multiple-output (MIMO) channels without requiring prior knowledge of channel state information (CSI).  We start by reviewing the conventional Kalman filtering-based state estimation algorithm, which typically relies on perfect CSI and incurs considerable computational complexity. To overcome the need for CSI, we introduce a novel semantic aggregation method, in which sensors transmit semantic measurement discrepancies to the remote state estimator through analog aggregation. To further reduce computational complexity, we introduce a constant-gain-based filtering algorithm that can be optimized offline using the constrained stochastic successive convex approximation (CSSCA) method. We derive a closed-form sufficient condition for the estimation stability of our proposed scheme via Lyapunov drift analysis. Numerical results showcase significant performance gains using the proposed scheme compared to several widely used methods.

\end{abstract}


\section{Introduction}
\begin{figure}
    \centering
    \includegraphics[height=4.9cm,width=9.3cm]{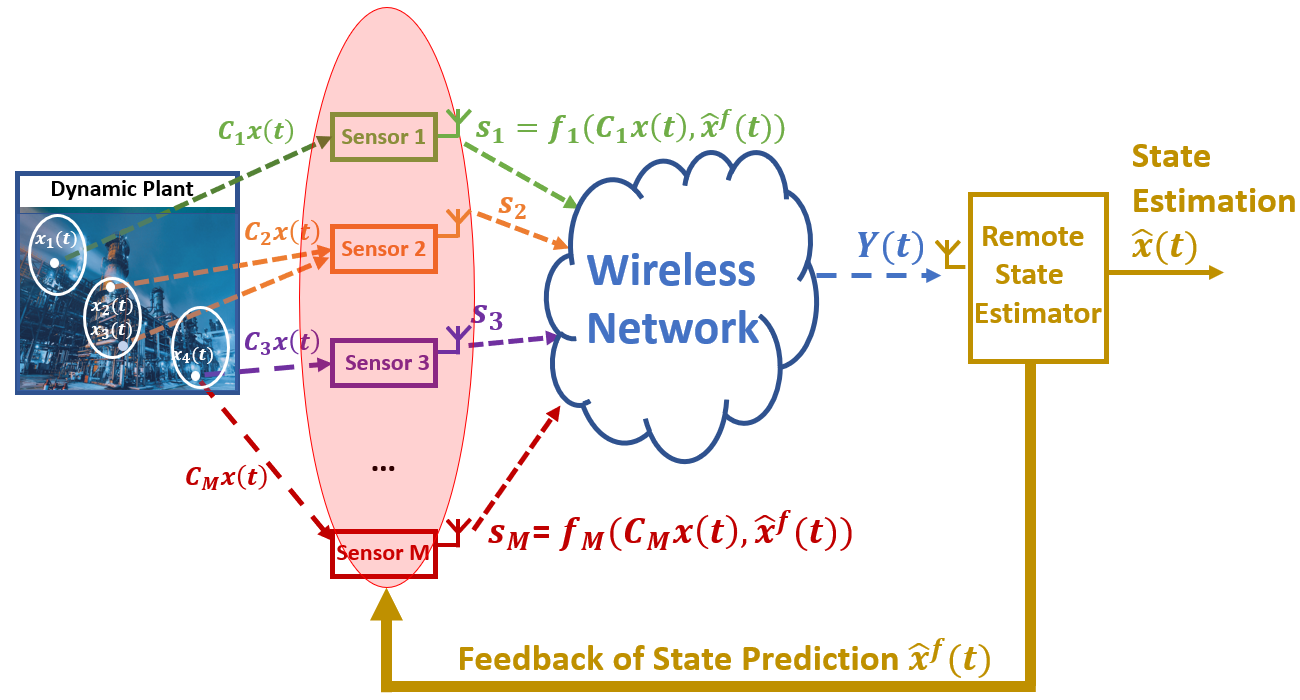}
    \caption{Typical architecture of a remote state estimation system over a wireless network.}
    \label{architecture}
\end{figure}

Remote state estimation has gained widespread interest recently \cite{ding2020secure}, with extensive applications in areas such as robotic control and autonomous navigation. A typical remote state estimation system consists of a physical \emph{dynamic plant}, multiple geographically scattered \emph{sensors}, and a \emph{remote state estimator}, as shown in Fig. \ref{architecture}. The wireless sensors measure the instantaneous internal states of the dynamic plant and deliver the state measurements to the remote state estimator over a wireless network. The remote state estimator generates the timely state estimation for the dynamic plant based on the received noisy measurements from the sensors. The wireless network between the remote state estimator and the sensors induces various impairments, such as fading and noise, which significantly degrade the estimation performance of the dynamic system.

Remote state estimation over a wireless network presents significant challenges. First, many existing works on remote state estimation overly simplify the wireless communication channels within the systems.
For example, \cite{xu2019event} utilizes an additive white Gaussian noise (AWGN) channel model, while \cite{peng2017optimal,ding2017multi} explores independent and identically distributed (i.i.d.) on-off failures in the wireless networks. However, these models overlook practical channel impairments, and directly applying such solutions to general fading channels may lead to estimation instability.

Second, most existing works on remote state estimation assume dedicated radio resource allocation for multiple sensors \cite{li2017detection}, resulting in poor spectral efficiency as sensor numbers grow. To address this, shared radio resource allocation for geographically dispersed sensors is preferred. Coordination of transmissions can be achieved using multiple access schemes such as ALOHA \cite{gao2022aloha} and non-orthogonal multiple access (NOMA) \cite{li2021joint}. However, these schemes face challenges with large sensor networks due to collision bottlenecks, and error propagation in the successive interference cancellation (SIC) process at the remote state estimator. 
Analog aggregation protocols \cite{tang2020remote}, which enable simultaneous transmission of sensor measurements without coordination, have recently been proposed to enhance spectral efficiency. However, these protocols rely on accurate CSI for state estimation, requiring pilot signal transmissions from each sensor for channel estimation at the remote state estimator \cite{liu2014channel}. This significantly increases communication overhead and introduces channel estimation noise \cite{liu2021deep}, which can undermine system stability.

Recent advancements in semantic communications, as highlighted in\cite{kountouris2021semantics,luo2022semantic,stavrou:2024}, emphasize the importance of transmitting relevant and useful application-specific information to reduce communication overhead. In the context of remote state estimation, this involves sending tailored sensor data rather than raw measurements. For instance, \cite{talebi2019distributed} investigates local Kalman filtering with central consensus, where local estimates are transmitted instead of raw data, potentially degrading performance due to measurement correlations. 
This underscores the need for further exploration of semantic communication strategies in remote state estimation.

In this work, we introduce a novel CSI-free, low-complexity remote state estimation algorithm over wireless MIMO fading channels using semantic analog aggregation. The contributions of this work are summarized as follows:
\begin{itemize}
    \item \textbf{Semantic Aggregation over Wireless MIMO Fading Channels.} Analog aggregation \cite{tang2020remote} improves spectral efficiency and reduces latency in mission-critical applications but incurs substantial overhead due to the transmission of raw sensor data and pilot signals. To address this, we propose a \textit{semantic aggregation scheme} in which sensors transmit only the discrepancy between predicted and actual measurements, leveraging predictions fed back from the remote state estimator. This approach preserves information integrity while drastically reducing communication overhead.

    \item \textbf{CSI-Free Low-Complexity State Estimation Solution.} 
Existing remote state estimation methods that rely on Kalman filtering \cite{peng2017optimal,ding2017multi,li2017detection,tang2020remote} require complex computations and perfect CSI. In contrast, we propose a constant-gain filtering algorithm that simplifies computation and eliminates the need for CSI. This approach avoids the transmission of pilot signals from the sensors and bypasses channel estimation at the remote state estimator, streamlining the entire process.

    \item  \textbf{Estimation Stability Analysis.}  
Using Lyapunov stability theory \cite{neely2022stochastic}, we derive a closed-form sufficient condition for ensuring estimation stability and propose a CSSCA-based offline algorithm to efficiently optimize the constant filtering gain.
\end{itemize}

\emph{Notation: }Uppercase and lowercase boldface letters denote matrices and vectors, respectively. The operators $(\cdot)^T$ and $\Tr(\cdot)$ represent the transpose and trace of a matrix, respectively. $\mathbf{0}_{m \times n}$ and $\mathbf{0}_m$ denote an $m \times n$ matrix and an $m \times m$ matrix with all elements equal to $0$, respectively. $\mathbf{I}_S$ represents the $S \times S$ identity matrix.
$\mathbb{R}^{m \times n}$, $\mathbb{S}_+^m$, $\mathbb{S}^m$, $\mathbb{Z}_+$, and $\mathbb{R}_+$ denote the set of $m \times n$ real matrices, the set of $m \times m$ positive definite matrices, the set of $m \times m$ positive semi-definite matrices, the set of positive integers, and the set of positive real numbers, respectively.
$||\mathbf{A}||$, $\mathbf{A}\|_F$ and $||\mathbf{a}||$ represent the spectral norm of a matrix $\mathbf{A}$, the Frobenius norm of a matrix $\mathbf{A}$ and the Euclidean norm of a vector $\mathbf{a}$, respectively. $[\mathbf{A}]_{i,j}$  denotes the $i$-th row and $j$-th column of the matrix $\mathbf{A}$. $[\mathbf{a}]_i$  denotes the $i$-th element of the vector $\mathbf{a}$.

\section{System Model}
In this section, we introduce the system model for the dynamic plant, wireless sensors, wireless communication channels, and remote state estimator.

\subsection{Dynamic Plant Model}
We consider a discrete-time system with $S\in\mathbb{Z}_+$ state variables. The physical plant state $\mathbf{x}(t)\in\mathbb{R}^{S\times 1}$ evolves according to first-order coupled linear difference equations:
\begin{equation}    \label{dynamic model}
\mathbf{x}(t+1)=\mathbf{A}\mathbf{x}(t)+\mathbf{w}(t),\,\,\,\,t=0,1,2,\ldots,
\end{equation}
where $\mathbf{A}\in\mathbb{R}^{S\times S}$ is the plant dynamics and  $\mathbf{w}(t)\sim\mathcal{N}(\mathbf{0}_{S\times 1}, \mathbf{W})$ is the plant noise with zero mean and finite noise covariance matrix $\mathbf{W}\in\mathbb{S}^{S}_+$. The initial plant state value $\mathbf{x}(0)=\mathbf{x}_0\in\mathbb{R}^{S\times 1}$.

\subsection{Wireless Sensor Model}
We consider a system consisting of $M\in\mathbb{Z}_+$ wireless sensors jointly monitoring the real-time plant state $\mathbf{x}(t)$. Each sensor has $N_t\in\mathbb{Z}_+$ transmission antennas and shares radio resources for high spectral efficiency.  The \emph{connection topology} of the wireless sensors defines the connectivity between the plant and sensors.
\newtheorem{definition}{Definition}
\begin{definition}
\emph{(Connection Topology of Sensors)}
The connection topology of (wireless) sensors is defined by a set of real matrices, $\left\{\mathbf{C}_1, \mathbf{C}_2,\ldots, \mathbf{C}_M\right\} \in \mathbb{R} ^{N_{t} \times S}$, where each $\mathbf{C}_m$ represents the connection between the dynamic plant and the $m$-th  sensor. Specifically, an element
$[\mathbf{C}_m]_{i,s}$ is non-zero if and only if the $i$-th element $[\mathbf{C}_{m}\mathbf{x}(t)]_i$  of the $m$-th sensor measurement 
includes $s$-th plant state component $[\mathbf{x}(t)]_i$. 
\end{definition}

\subsection{Semantic Signal Extraction and Wireless Communication Channel Model}
As illustrated in Fig. \ref{architecture}, each $m$-th  sensor accesses the plant state $\mathbf{x}(t)$ and converts it into an analog state measurement $\mathbf{C}_m\mathbf{x}(t)\in\mathbb{R}^{N_{t}\times 1}$. The sensor then generates the semantic signal $\mathbf{s}_m(t)\in\mathbb{R}^{N_t\times 1}$ by utilizing the state measurement $\mathbf{C}_m\mathbf{x}(t)$ and incorporating the real-time state prediction $\widehat{\mathbf{x}}^f(t)\in\mathbb{R}^{S\times 1}$ fed back from the remote state estimator, as follows.
\begin{equation}
    \mathbf{s}_m(t)=f_m(\mathbf{C}_m\mathbf{x}(t),\widehat{\mathbf{x}}^f(t))
\end{equation}
where $f_m(\cdot)$ represents the semantic signal extraction process for the $m$-th sensor and will be detailed further in Section \ref{section:semantic aggregation}.

The semantic signals $\left\{\mathbf{s}_1(t),\ldots,\mathbf{s}_M(t)\right\}$ from $M$ sensors are delivered to the remote state estimator  
through analog aggregation\cite{tang2020remote}. 
 The received signal $\mathbf{y}(t)\in\mathbb{R}^{N_{r\times 1}}$ at the remote state estimator represents the aggregated semantic signals from multiple sensors, given by\footnote{In this work, we assume the active signals $\left\{\mathbf{s}_m(t)\right\}$ are synchronized in transmission, which can be achieved using traditional methods like pilot-aided or decision-directed synchronization \cite{moretti2013combined}. }
\begin{equation}\label{analog-aggregation}
    \mathbf{y}(t)=\sum_{m=1}^M \delta_{m}(t)\mathbf{H}_m(t)\mathbf{s}_m(t)+\mathbf{v}(t),
\end{equation}
where $N_r\in\mathbb{Z}_+$ is the number of receiving antennas at the remote state estimator. $\delta_m(t)\in\left\{0, 1\right\}$ models the random activity of the $m$-th sensor, which is i.i.d. over the sensors and timeslots, satisfying $\Pr(\delta_m(t)=1)=p\in[0, 1]$.
$\mathbf{H}_m(t)\in\mathbb{R}^{N_r\times N_t}$ represents the MIMO channel gain, capturing path loss, shadowing, and fading between the 
$m$-th sensor and the remote state estimator. \footnote{To simplify the notation, we assume the system is real. The model can be
extended to a complex case by augmenting the complex symbol into $\mathbb{R}^2$.}  It remains constant within each timeslot and is i.i.d. over sensors and timeslots. Each element of $\mathbf{H}_m(t)$ follows a certain distribution with mean $h\in\mathbb{R}$, and finite variance $\sigma^2>0$. $\mathbf{v}(t)\sim\mathcal{N}(\mathbf{0}_{N_r\times 1},\mathbf{I}_{N_{r}})$ is the additive channel noise at the remote state estimator. 

\emph{Remark: (Modulation-Free Analog Aggregation \cite{tang2020remote})} Conventional digital transmission focuses on decoding individual signals using modulation to differentiate devices. In contrast, our approach prioritizes state estimation by leveraging collided, modulation-free analog sensor measurements, which inherently capture richer information about the plant state, leading to significantly improved estimation performance.

\subsection{State Estimation at the Remote State Estimator}
The objective of the remote state estimator is to compute the minimum mean square error (MMSE) estimate of the real-time plant state $\mathbf{x}(t)$  based on the received signal $\mathbf{y}_0^t\triangleq\left\{\mathbf{y}(0),\ldots,\mathbf{y}(t)\right\}$ from the sensors. For brevity, we denote:
\begin{equation}
    \widehat{\mathbf{x}}(t) \triangleq \mathbb{E}[\mathbf{x}(t)|\mathbf{y}_0^t];
\end{equation}
\begin{equation}
    \widehat{\mathbf{x}}^f(t)\triangleq \mathbb{E}[\mathbf{x}(t)|\mathbf{y}_0^{t-1}];
\end{equation}
\begin{equation}
   \mathbf{P}(t)\triangleq\mathbb{E} [(\mathbf{x}(t)- \widehat{\mathbf{x}}(t))(\mathbf{x}(t)- \widehat{\mathbf{x}}(t))^{T}|\mathbf{y}_0^t];
\end{equation}
\begin{equation}
 \!\!\!\!\!\mathbf{P}^f(t) =\mathbb{E} [(\mathbf{x}(t)- \widehat{\mathbf{x}}^f(t))(\mathbf{x}(t)- \widehat{\mathbf{x}}^f(t))^{T}|\mathbf{y}_0^{t-1}],
\end{equation}
where $\widehat{\mathbf{x}}(t)\in\mathbb{R}^{S\times 1}$, $\widehat{\mathbf{x}}^f(t)\in\mathbb{R}^{S\times 1}$, $\mathbf{P}(t)\in\mathbb{S}_+^S$ and $\mathbf{P}^f(t)\in\mathbb{S}_+^S$
denote the estimated plant state, predicted plant state, posterior error covariance matrix, and prior error covariance matrix for the plant state at $t$-th timeslot, respectively.

In traditional Kalman filtering, the signal $\mathbf{y}(t)$ at the remote state estimator is an aggregation of raw sensor measurements, given by $\mathbf{s}_m(t)=\mathbf{C}_m\mathbf{x}(t)$. The state estimation algorithm is divided into two sequential steps: the \emph{prediction step} and the \emph{estimation step}, both performed in an online manner. Specifically, during the prediction step, the predicted plant state $\widehat{\mathbf{x}}^f(t)$ and the prior error covariance matrix $\mathbf{P}^f(t)$ are updated as follows:
\begin{equation}\label{prediction-equation}
    \widehat{\mathbf{x}}^f(t)=\mathbf{A} \widehat{\mathbf{x}}(t-1);
\end{equation}
\begin{equation}
    \mathbf{P}^f(t)=\mathbf{A}\mathbf{P}(t-1)\mathbf{A}^{T}+\mathbf{W}.
\end{equation}
In the estimation step, the estimated plant state $\widehat{\mathbf{x}}(t)$ and the posterior error covariance matrix $\mathbf{P}(t)$ are updated as follows:
\begin{equation}\label{innovation}
    \widehat{\mathbf{x}}(t)=    \underbrace{\widehat{\mathbf{x}}^f(t)}_{predicted\,value}+{\mathbf{K}}(t)\underbrace{\Pi(t)}_{innovative\,deviation};
\end{equation}
\begin{equation}\label{error-covariance}
\begin{split}
&\!\!\!\!\!\!\!\!\mathbf{P}(t)=(\mathbf{I}_S-{\mathbf{K}}(t)\mathbf{H}(t))\mathbf{P}^f(t)(\mathbf{I}_S-{\mathbf{K}}(t)\mathbf{H}(t))^{T}+{\mathbf{K}}(t)\mathbf{K}^T(t),
\end{split}
\end{equation}
where $\Pi(t)=(\mathbf{y}(t)-\mathbf{H}(t)\widehat{\mathbf{x}}^f(t))\in\mathbb{R}^{N_r\times 1}$,
$\mathbf{H}(t)=\sum_{m=1}^M\delta_m(t)\mathbf{H}_m(t)\mathbf{C}_m\in\mathbb{R}^{N_r\times S}$ represents the CSI and
$\mathbf{K}(t)\in\mathbb{R}^{S\times N_{r}}$ denotes the Kalman filtering gain, given by
\begin{equation}\label{kalman filter}
\begin{split}
    &\mathbf{K}(t)=\mathbf{P}^f(t)\mathbf{H}^T(t)(\mathbf{H}(t)\mathbf{P}^f(t)\mathbf{H}^T(t)+\mathbf{I}_{N_r})^{-1}.
\end{split}
\end{equation}

The Kalman filtering equations (\ref{prediction-equation})-(\ref{kalman filter}) present several challenges that must be carefully addressed:
\begin{itemize}
    \item \textbf{High Computational Complexity.} Matrix inversions in Kalman filtering involve a computational cost of
 $\mathcal{O}(S^3)$, which becomes significant for large $S$.
 To reduce complexity, \cite{katewa2020minimum} proposed a fixed-gain estimator for static channels (i.e., $\mathbf{H}(t)=\mathbf{H}\in\mathbb{R}^{N_r\times N_t}, \mathbf{K}(t)=\mathbf{K}\in\mathbb{R}^{S\times N_r}$) using pole-zero placement to satisfy the stability condition $\|\mathbf{A}-\mathbf{A}\mathbf{K}\mathbf{H}\|<1$.
However, this method does not apply to our time-varying system, where $\mathbf{H}(t)$ changes dynamically, as continuously maintaining the stability condition under such variations is highly challenging.

\item \textbf{Perfect CSI Requirement.} As indicated in (\ref{innovation}) and (\ref{error-covariance}), Kalman filtering requires precise knowledge of the CSI, denoted as $\mathbf{H}(t)$. To acquire $\mathbf{H}(t)$, channel estimation has to be performed at the remote state estimator, which relies on transmitted pilot symbols from all sensors. However, this procedure incurs considerable communication overhead at the sensors due to pilot signal transmissions. Moreover, channel estimation is prone to noise \cite{liu2014channel}, which can further degrade performance. Applying Kalman filtering with inaccurate CSI through brute-force methods can severely compromise the quality of state estimation.
\end{itemize}

In the following section, we propose a CSI-free, low-complexity state estimation solution that involves semantic information delivery at sensors and static gain-based filtering at the remote state estimator. We demonstrate that estimation stability can be maintained in our time-varying system by appropriately designing the filtering gain through an offline optimization process.

\section{CSI-Free Low-Complexity Remote State Estimation via Semantic Analog Aggregation\label{section:semantic aggregation}}
In this section, we propose a semantics-empowered remote state estimation scheme. We present the process of semantic information generation at the sensors and a state estimation algorithm at the remote estimator with a static filtering gain. Through Lyapunov drift analysis, we derive a stability condition and propose an offline CSSCA algorithm to optimize the filtering gain.

\subsection{Generation of Semantic Information  at Sensors}
In conventional Kalman filtering, sensor measurements are used to calculate the innovative deviation during the estimation step, as shown in (\ref{innovation}). Consequently, to optimize communication, instead of transmitting the raw sensor measurement $\mathbf{C}_m\mathbf{x}(t)$, we can exploit the structure of Kalman filtering by transmitting only the discrepancy between the predicted and actual measurements at each $m$-th sensor, given by
$\mathbf{C}_m\mathbf{x}(t)-\mathbf{C}_m\widehat{\mathbf{x}}^f(t)$, where the state prediction $\widehat{\mathbf{x}}^f(t)$ is fed back from the remote state estimator at the $(t-1)$-th timeslot. In mathematical terms, this can be expressed as:
 \begin{equation}
    \label{extraction}\mathbf{s}_m(t)=\mathbf{C}_m(\mathbf{x}(t)-\widehat{\mathbf{x}}^f(t)).
\end{equation}
Our design ensures that the aggregated signal $\mathbf{y}(t)$ at the remote state estimator equals the innovation $\Pi(t)$ as defined in (\ref{innovation}). Consequently, the estimator effectively captures the Kalman filtering innovation without requiring CSI  $\mathbf{H}(t)$.

It is also noteworthy that the discrepancy between the predicted and the actual measurements, denoted as $\mathbf{C}_m\mathbf{x}(t)-\mathbf{C}_m\widehat{\mathbf{x}}^f(t)$, is generally smaller in magnitude than the actual state measurement $\mathbf{C}_m\mathbf{x}(t)$, i.e., $\|\mathbf{C}_m\mathbf{x}(t)-\mathbf{C}_m\widehat{\mathbf{x}}^f(t)\|<<\|\mathbf{C}_m\mathbf{x}(t)\|$. As a result, our proposed method not only improves estimation efficiency but also significantly reduces power consumption at the wireless sensors.

\subsection{CSI-Free Low-Complexity State Estimation Algorithm at the Remote State Estimator}
By leveraging a constant filtering gain at the remote state estimator, we can implement a CSI-free, low-complexity state estimation algorithm that builds upon traditional Kalman filtering principles, as detailed in Algorithm \ref{algorithm}. Removing the need for CSI offers significant advantages for both the sensors and the remote state estimator. From the perspective of the sensors, this approach significantly reduces the communication overhead caused by pilot transmissions. On the side of the remote state estimator, it removes the necessity of channel estimation, thereby reducing computational complexity and also circumventing the degraded state estimation performance typically caused by channel estimation noise in existing approaches (e.g., see \cite{liu2021deep}).

\begin{algorithm}
\small
\caption{CSI-Free Low-Complexity Remote State Estimation via Semantic Analog Aggregation} 

\textbf{Initialization:} $\widehat{\mathbf{x}}^f(0)\sim\mathcal{N}(\mathbf{0}_{S\times 1},\mathbf{1}_S)$.

\textbf{For $t=0,1,...$:}

\begin{itemize}
    \item \textbf{Step 1: (Information Collection at Sensors)}\\
    Each $m$-th sensor  obtains $\mathbf{C}_m\mathbf{x}(t)$ and  stores $\widehat{\mathbf{x}}^f(t)$  fed back from the remote state estimator.

    \item  \textbf{Step 2: (Signal Extraction and Transmission at Sensors)}
        \\$\mathbf{s}_m(t) \leftarrow$ Compute (13) using $\mathbf{C}_m\mathbf{x}(t)$ and   $\widehat{\mathbf{x}}^f(t)$; 
    \\ $\mathbf{y}(t) \leftarrow$ Compute (3) using $\mathbf{s}_m(t)$.
    \item \textbf{Step 3: (State Estimation at the Remote State Estimator)} 
    \begin{equation}\label{innovation-proprosed}
\widehat{\mathbf{x}}(t)\leftarrow\widehat{\mathbf{x}}^f(t)+\mathbf{K}\mathbf{y}(t),
\end{equation} where $\mathbf{K}\in\mathbb{R}^{S\times N_{r}}$ is the constant filtering gain to be designed offline as described in Section \ref{subsection:gain-optimization}.
    \item  \textbf{Step 4: (State Prediction at the Remote State Estimator)}
    
    \begin{equation}\label{prediction-proposed}
\widehat{\mathbf{x}}^f(t+1)\leftarrow\mathbf{A}\widehat{\mathbf{x}}(t).
\end{equation}
\\The predicted state $\widehat{\mathbf{x}}^f(t+1)$ is broadcast to the sensors. 
\end{itemize}

\label{algorithm}
\end{algorithm}

\subsection{Estimation Stability Analysis}
We employ Lyapunov stability theory \cite{neely2022stochastic} to establish the sufficient condition for estimation stability, i.e., $\limsup_{T\rightarrow\infty}\frac{1}{T}\sum_{t=1}^T\mathbb{E}[\Tr(\mathbf{P}^f(t))]<\infty$, via our proposed scheme for wireless MIMO fading channels. To facilitate this analysis, we define a Lyapunov function as follows:
\begin{equation}\label{lyapunov-function}
    \mathcal{L}(\mathbf{P}^f(t))=\Tr(\mathbf{P}^f(t)),
\end{equation}
and the associated Lyapunov drift is given by
\begin{equation}\label{lyapunov-drift}
    \Gamma(\mathbf{P}^f(t))=\mathbb{E}[\mathcal{L}(\mathbf{P}^f(t+1))-\mathcal{L}(\mathbf{P}^f(t))|\mathcal{L}(\mathbf{P}^f(t))].
\end{equation}
Substituting (\ref{innovation-proprosed}), (\ref{prediction-proposed}) and (\ref{lyapunov-function}) into (\ref{lyapunov-drift}), we have the following theorem on the Lyapunov drift.
\begin{theorem}\emph{(Lyapunov Drift)} The Lyapunov drift $\Gamma(\mathbf{P}(t))$ is bounded as:
\begin{equation}\nonumber
    \Gamma(\mathbf{P}^f(t))\leq N_r\|\mathbf{A}\|^2\|\mathbf{K}\|^2+\Tr(\mathbf{W})+\mathbb{E}[\Tr( (\mathbf{A}-\mathbf{A}\mathbf{K}\mathbf{H}(t))^T
\end{equation}
\begin{equation}\label{lyapunov-drift-bound}
    (\mathbf{A}-\mathbf{A}\mathbf{K}\mathbf{H}(t)))]\Tr(\mathbf{P}^f(t))-\Tr(\mathbf{P}^f(t)).
\end{equation}

\emph{Proof:} See Appendix A.

\end{theorem}

The sufficient condition for estimation stability can be determined by analyzing whether the Lyapunov drift in (\ref{lyapunov-drift-bound}) is negative, as summarized in the following theorem.
\begin{theorem}\emph{(Sufficient Condition for Estimation Stability)} If $\mathbb{E}[\| \mathbf{I}_S-\mathbf{K}\mathbf{H}(t)\|^2_F]\leq (\|\mathbf{A}\|^2)^{-1}$,
then the system is stable with state estimation MSE upper bounded by:
\begin{equation}\label{mse-bound}
    \begin{split}
        &\limsup_{T\rightarrow\infty}\frac{1}{T}\mathbb{E}[\sum_{t=1}^T\Tr(\mathbf{P}^f(t))]<\frac{\Tr(\mathbf{W})+N_r\|\mathbf{A}\|^2\|\mathbf{K}\|^2}{1-\|\mathbf{A}\|^2 \mathbb{E}[\| \mathbf{I}_S-\mathbf{K}\mathbf{H}(t)\|^2_F]}.
    \end{split}
\end{equation}

\emph{Proof:} See Appendix B.
\label{mse-bound-theorem}
\end{theorem}

\subsection{Optimization of Filtering Gain\label{subsection:gain-optimization}}
The objective of remote state estimation is to minimize the state estimation MSE while ensuring estimation stability. As a result, the constant filtering gain $\mathbf{K}$ should be designed to minimize the right-hand side (R.H.S.) of (\ref{mse-bound}), in line with the sufficient condition for estimation stability outlined in Theorem \ref{mse-bound-theorem}. This leads to the following optimization problem for 
 $\mathbf{K}$.

\emph{Problem 1: (Constant Filtering Gain Optimization)} The optimal gain $\mathbf{K}^*$ can be obtained by solving the following optimization problem:
\begin{equation}
    \begin{split}
        \max_{\mathbf{K}} f_0(\mathbf{K}),\,\,\,\, \text{s.t.,}\,\,\, f_1(\mathbf{K})>0,
    \end{split}
\end{equation}
where $f_0(\mathbf{K})=\frac{1-\|\mathbf{A}\|^2 \mathbb{E}[\| \mathbf{I}_S-\mathbf{K}\mathbf{H}(t)\|^2_F]}{\Tr(\mathbf{W})+N_r\|\mathbf{A}\|^2\|\mathbf{K}\|^2}$ and $f_1(\mathbf{K})=\frac{1}{\|\mathbf{A}\|^2}- \mathbb{E}[\| \mathbf{I}_S-\mathbf{K}\mathbf{H}(t)\|^2_F]$.

Note that Problem $1$ is a non-convex optimization problem. Our goal is to design an efficient offline CSSCA algorithm \cite{liu2019stochastic} to identify a stationary point. Specifically, in each $r$-th iteration, we generate a synthetic $\mathbf{H}^r(t)$ according to the distribution of $\mathbf{H}(t)$. Subsequently, $\mathbf{K}^{r+1}$ is obtained by solving the following problem.

\emph{Problem 2: (Iterative Convex Surrogate Approximation \cite{liu2019stochastic})}
\begin{equation}
\label{problem-iterative-opt}
    \max_{\mathbf{K}} \bar{f}_0^r(\mathbf{K}),\,\,\,\,\bar{f}_1^r(\mathbf{K})>0,
\end{equation}
where
\begin{align}
\label{objective-surrogate}
   &\bar{f}_0^r(\mathbf{K})=(1-\tau_0^r)\bar{f}_0^{r-1}(\mathbf{K})+\tau_0^r(f_0(\mathbf{K}^r)+\nabla^T_{\mathbf{K}^r}f_0( \mathbf{K}^r )\nonumber\\
   &(\mathbf{K}-\mathbf{K}^r)+\epsilon_0\|\mathbf{K}-\mathbf{K}^r\|^2),
\end{align}
with

\begin{align}
\label{constraint-surrogate}
   &\bar{f}_1^r(\mathbf{K})=(1-\tau_1^r)\bar{f}_1^{r-1}(\mathbf{K})+\tau_1^r(f_1(\mathbf{K}^r)+\nabla^T_{(\mathbf{K})^r}f_1( \mathbf{K}^r )\nonumber\\
   &(\mathbf{K}-\mathbf{K}^r)+\epsilon_1\|\mathbf{K}-\mathbf{K}^r\|^2).
\end{align}
Note that $\tau_0,\tau_1\in (0,1]$ are chosen via the Armijo step-size rule. $\epsilon_0,\epsilon_1<0$ can be any constants.

We summarize the offline optimization algorithm for $\mathbf{K}$ in Algorithm 2 and conclude this section with a lemma on the convergence of Algorithm 2.

\begin{algorithm}[H]
\small
\caption{Offline Optimization of Constant Filtering Gain} 
\textbf{Initialization:} The total iteration time $R$; Initial the constant gain \,\,\, $\mathbf{K}^0\in\mathbb{R}^{S\times N_r}$.

\textbf{For $r=1,2,...,R$}:

\begin{itemize}
    \item \textbf{Step 1: (Update of the Surrogate)} \\ $\bar{f}_0(\mathbf{K})\leftarrow$ Using (\ref{objective-surrogate}),  $\mathbf{H}^r(t)$ and $\mathbf{K}^r$;\\ $\bar{f}_1(\mathbf{K})\leftarrow$ Using (\ref{constraint-surrogate}),  $\mathbf{H}^r(t)$ and $\mathbf{K}^r$.
    \item \textbf{Step 2: (Update of the Gain)}\\ $\mathbf{K}^{r+1}\leftarrow$ solving (\ref{problem-iterative-opt}).
\end{itemize}

\textbf{Output:} Obtaining  solution $\mathbf{K}^{R}$ to Problem $1$.

\end{algorithm}

\begin{lemma}
\emph{(Convergence of Algorithm 2\cite{liu2019stochastic})} Algorithm $2$ converges to the stationary point $\mathbf{K}^*$ of Problem $1$ almost surely, i.e., $\Pr(\lim_{R\rightarrow\infty}\mathbf{K}^R=\mathbf{K}^*)=1$.
\end{lemma}

\section{Numerical Results}
In this section, we assess the performance of the proposed remote state estimation algorithm by comparing it against the following baseline schemes:
\begin{itemize}
\item \textbf{Baseline 1}: \emph{(Kalman Filtering under ALOHA)} Each sensor activates to transmit raw measurement with a fixed probability $p_a\in[0, 1]$ per timeslot. In the event of simultaneous activations, collision resolution is performed at the remote estimator. Each active sensor transmits a dedicated pilot $\mathbf{T}_m\in\mathbb{R}^{N_t\times N_t}$. The remote estimator uses Kalman filtering, with CSI estimated using least-squares-based channel estimation from the pilots.
    
\item \textbf{Baseline 2}: \emph{(Kalman Filtering under Random TDMA)} The remote estimator randomly selects one sensor to transmit its raw measurement at each timeslot. The state estimation method used is identical to that of Baseline 1.

\item \textbf{Baseline 3}: \emph{(Kalman Filtering under Analog Aggregation)} Each sensor that carries raw measurements randomly accesses the wireless network with a fixed probability $p\in [0, 1]$ using analog aggregation. The state estimation method remains the same as in Baseline 1.
\end{itemize}

We model the system dynamics as:
\begin{equation}
    \mathbf{x}(t+1)=\begin{bmatrix}1.01&0.05&0.01\\0.02&0.98&0.01\\0.003&0.002&0.98\end{bmatrix}\mathbf{x}(t)+\mathbf{w}(t),
\end{equation}
where $\mathbf{w}(t)\sim\mathcal{N}(\mathbf{0}_{3\times 1},\mathbf{I}_{3})$. Each element of the MIMO channel fading $\mathbf{H}_m(t)\in\mathbb{R}^{3\times 3}$ is Rayleigh distributed with a scalar parameter of $3$, associated with the variance of the Gaussian components. The signal-to-noise (SNR) ratio is $12.5$ dB.
The sensors sequentially measure the plant state, with each $m$-th sensor observing the $((m-1)\mod 3 +1)$-th state. Specifically, the matrix $\mathbf{C}_m \in \mathbb{R}^{3\times 3}$ where $1\leq m\leq M,$ is defined such that $[\mathbf{C}_m]_{i,(m-1)\mod 3 +1}=0.1$ for $1\leq i\leq 3$, while all other elements are set to zero. Each sensor is activated randomly with a fixed probability of $p=p_a=0.3$. 

\begin{figure}
    \centering
    \includegraphics[height=4.1cm,width=7.6cm]{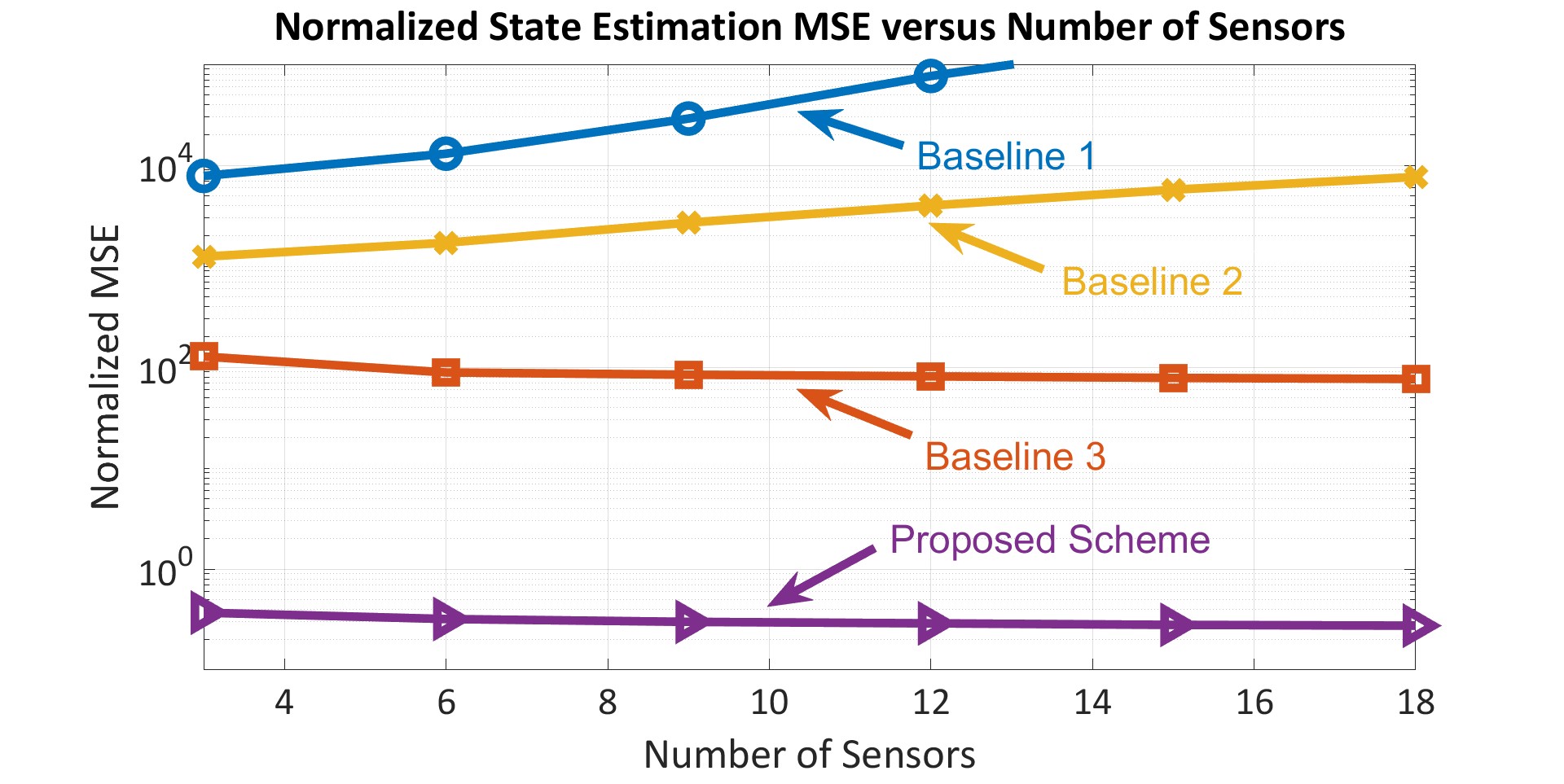}
    \caption{Normalized state estimation MSE versus the number of sensors $M$.}
    \label{sensor}
\end{figure}

\subsection{NMSE versus the Number of Sensors}
Fig. \ref{sensor} shows the normalized MSE (NMSE) of state estimation versus the number of sensors $M$. Baselines 1 and 2 exhibit degraded performance as the number of sensors increases, primarily due to collision issues in Baseline 1 and access latency bottlenecks in Baseline 2. In contrast, both Baseline 3 and our proposed scheme benefit from an increasing number of sensors through the use of analog aggregation. However, our proposed scheme outperforms Baseline 3 by achieving a smaller NMSE, as it avoids reliance on imperfect CSI for state estimation.

\subsection{Total Transmission Power at Sensors versus Timeslot}
\begin{figure}
    \centering
    \includegraphics[height=4.1cm,width=7.7cm]{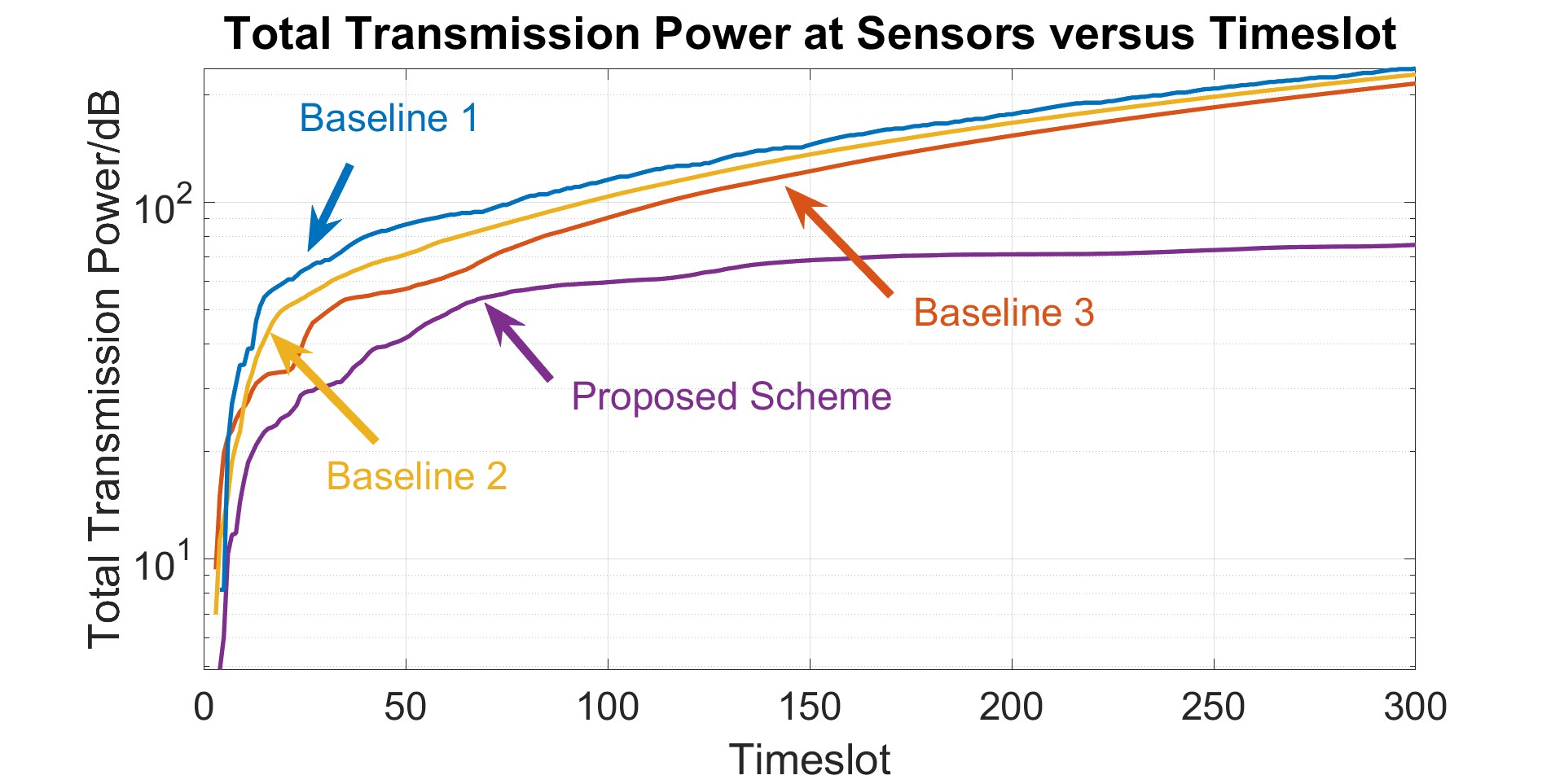}
    \caption{Total transmission power at sensors versus timeslot. The number of sensors $M=6$.}
    \label{power}
\end{figure}

Fig. \ref{power} illustrates the total transmission power at sensors $\sum_{m=1}^M\|\mathbf{s}_m(t)\|^2$ over time, highlighting the communication overhead associated with state estimation. Specifically, the figure illustrates that our proposed approach results in a substantial power savings of at least 40.4\% over 300 timeslots, compared to baseline schemes that transmit full state measurements and pilot symbols.

\subsection{Total CPU Computational Time versus Plant Dimension}
\begin{figure}
    \centering
    \includegraphics[height=4.1cm,width=7.6cm]{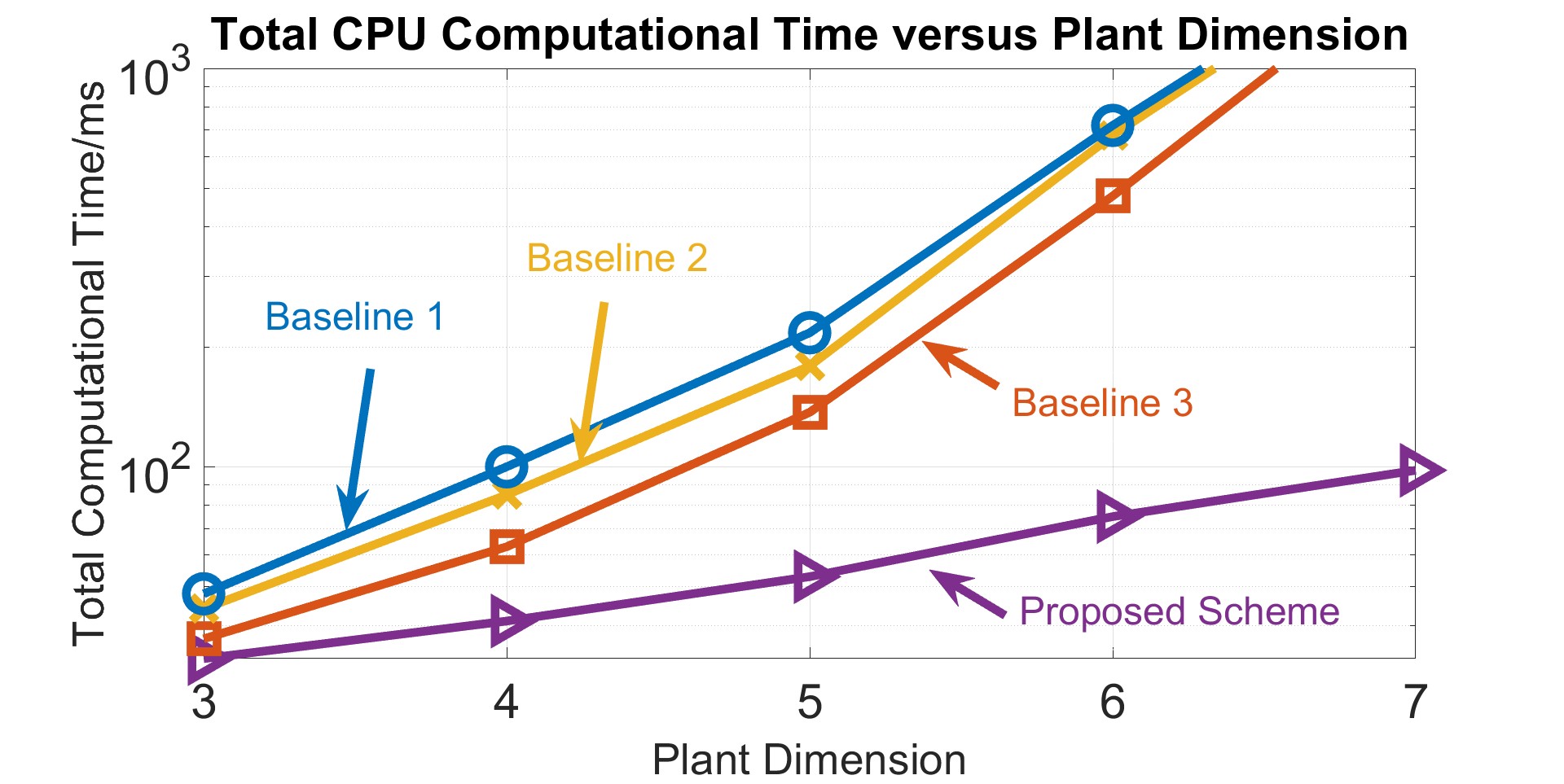}
    \caption{Total CPU computational time versus plant dimension. $\mathbf{A}\in\mathbb{R}^{S\times S}$ and $\mathbf{C}_m\in\mathbb{R}^{N_t\times S}$ are generated with elements following a Gaussian distribution (zero mean, unit variance). The number of sensors $M=6$.}
    \label{complexity}
\end{figure}

Fig. \ref{complexity} shows the total CPU computational time as a function of the plant dimension over $10^4$ timeslots. The results show that our proposed constant-gain-based filtering scheme reduces computational time by at least 23.9\% compared to baseline schemes, which rely on the more computationally intensive standard Kalman filtering.

\section{Conclusions}
In this paper, we addressed the challenge of CSI-free, low-complexity remote state estimation over wireless MIMO fading channels. We started by examining the limitations of the traditional Kalman-filtering-based state estimation algorithm, which requires accurate CSI and incurs substantial computational complexity. To overcome these issues, we proposed a novel semantic aggregation approach that consolidates semantic information at the remote state estimator, enabling static-gain-based filtering. Through Lyapunov drift analysis, we demonstrated that our approach stabilizes the system through careful offline design of the static filtering gain. Simulation results highlighted the superiority of our method compared to widely used existing techniques.

\appendix

\subsection{Proof of Theorem 1}
Let $\mathbf{P}(t)$ and $\mathbf{P}^f(t)$ represent the posterior and prior state estimation error covariances, respectively. The state prediction in \textbf{Step 4} of Algorithm 1 yields the following error covariance evolution:
\begin{equation}
    \mathbf{P}^f(t+1)=\mathbf{A}\mathbf{P}(t)\mathbf{A}^T+\mathbf{W}.
    \label{predicted error covariance}
\end{equation}
The state estimation via \textbf{Step 3} of Algorithm 1 gives the error covariance evolution as follows:
\begin{equation}
\begin{split}
   &\mathbf{P}(t)=(\mathbf{I}_S-\mathbf{K}\mathbf{H}(t))\mathbf{P}^f(t) (\mathbf{I}_S-\mathbf{K}\mathbf{H}(t))^T+\mathbf{K}\mathbf{K}^T.
   \label{estimated error covariance}
\end{split}
\end{equation}
Combining (\ref{predicted error covariance}) and (\ref{estimated error covariance}) leads to
\begin{equation}\nonumber
\begin{split}
    &\mathbf{P}^f(t+1)=\mathbf{A}(\mathbf{I}_S-\mathbf{K}\mathbf{H}(t))\mathbf{P}^f(t)(\mathbf{I}_S-\mathbf{K}\mathbf{H}(t))^T\mathbf{A}^T+
  \end{split}
\end{equation}  
\begin{equation}
\mathbf{A}\mathbf{K}\mathbf{K}^T\mathbf{A}^T+\mathbf{W}.
\label{overall}
\end{equation}
Substituting (\ref{overall}) into (\ref{lyapunov-drift}) yields
\begin{equation}
\begin{split}
&\Gamma(\mathbf{P}^f(t))=\mathbb{E}[\Tr( (\mathbf{A}-\mathbf{A}\mathbf{K}\mathbf{H}(t))\mathbf{P}^f(t)(\mathbf{A}-\mathbf{A}\mathbf{K}\\&\mathbf{H}(t))^T)|\mathbf{P}^f(t)]+ \mathbb{E}[\Tr(\mathbf{A}\mathbf{K}\mathbf{K}^T\mathbf{A}^T)]+\Tr(\mathbf{W}).
\label{lyapunov}
\end{split}
\end{equation}
Note that the R.H.S. of (\ref{lyapunov}) can be further upper bounded by the R.H.S. of (\ref{lyapunov-drift-bound}). This concludes the proof.

\subsection{Proof of  Theorem 2}
Note that $\limsup_{T\rightarrow\infty}\frac{1}{T}\mathbb{E}[\sum_{t=1}^T\mathcal{L}(\mathbf{P}^f(t))]<\infty$ if $\mathbb{E}[\mathcal{L}(\mathbf{P}^f(t+1))-\mathcal{L}(\mathbf{P}(t))|\mathcal{L}(\mathbf{P}^f(t))]<\alpha(t)\mathcal{L}(\mathbf{P}^f(t))+\beta(t)$, where $\alpha(t)<0$ and $\beta(t)>0, \forall t\geq 0$. As a result, the system is stable if $\mathbb{E}[\| \mathbf{I}_S-\mathbf{K}\mathbf{H}(t)\|^2_F]\leq (\|\mathbf{A}\|^2)^{-1}$.

Taking the expectation over random $\mathbf{P}^f(t)$ on both sides of (\ref{lyapunov-drift-bound}), it gives that
\begin{equation}
\begin{split}
&\mathbb{E}[\Tr(\mathbf{P}^f(t+1))]-\mathbb{E}[\Tr(\mathbf{P}^f(t))]\leq N_r\|\mathbf{A}\|^2\|\mathbf{K}\|^2+\Tr(\mathbf{W})\\&
+\|\mathbf{A}\|^2\mathbb{E}[\|(\mathbf{I}_S-\mathbf{K}\mathbf{H}(t))(\mathbf{I}_S-\mathbf{K}\mathbf{H}(t))^T\|-1]\mathbb{E}[\Tr(\mathbf{P}^f(t))].
\label{average bound}
\end{split}
\end{equation}
Summing up (\ref{average bound}) over the timeslot $t\in\left\{0,2,\ldots,T\right\}$ yields a \emph{telescoping series}.
Subsequently, dividing the induced inequality over $T$, using the fact that $\Tr(\mathbf{P}(T+1))\geq 0$, and
taking the limit over $T$,  gives (\ref{mse-bound}). This concludes the proof.

\section*{Acknowledgment}
This work is supported by the SNS JU project 6G-GOALS \cite{strinati:2024} under the EU’s Horizon programme Grant Agreement No. 101139232. The work of P. A. Stavrou is also supported in part by the Huawei France-EURECOM Chair on Future Wireless Networks.

\bibliographystyle{IEEEtran}
\bibliography{IEEEabrv,Bibliography}
\end{document}